\documentclass[preprint,showpacs,preprintnumbers,amsmath,amssymb]{revtex4}

\usepackage{graphicx}
\usepackage{dcolumn}
\usepackage{bm}
\usepackage{amssymb}

\begin{document}
\title{
Classical open systems with nonlinear nonlocal dissipation 
and state-dependent diffusion: Dynamical responses \\ 
and the Jarzynski equality 
}

\author{Hideo Hasegawa}
\altaffiliation{hideohasegawa@goo.jp}
\affiliation{Department of Physics, Tokyo Gakugei University,  
Koganei, Tokyo 184-8501, Japan}%

\date{\today}
\begin{abstract}
We have studied dynamical responses and the Jarzynski equality (JE) of classical open 
systems described by the generalized Caldeira-Leggett model with the nonlocal
system-bath coupling. In the derived non-Markovian Langevin equation,
the nonlinear nonlocal dissipative term and state-dependent diffusion term yielding 
multiplicative colored noise satisfy the fluctuation-dissipation relation.
Simulation results for harmonic oscillator systems have shown the following:
(a) averaged responses of the system $\langle x \rangle$ to applied sinusoidal
and step forces significantly depend on model parameters of magnitudes of additive 
and multiplicative noises and the relaxation time of colored noise, 
although stationary marginal probability distribution functions are independent of them, 
(b) a combined effect of nonlinear dissipation and multiplicative colored noise induces
enhanced fluctuations $\langle [x-\langle x\rangle]^2 \rangle$ for an applied sinusoidal 
force, and (c) the JE holds for an applied ramp force
independently of the model parameters with a work distribution function 
which is (symmetric) Gaussian and asymmetric non-Gaussian for additive and multiplicative 
noises, respectively.
It has been shown that the non-Markovian Langevin equation in the local and over-damped limits 
is quite different from the widely adopted phenomenological Markovian Langevin equation 
subjected to multiplicative noise.

\end{abstract}

        

\maketitle
\newpage
\section{Introduction}

In the last almost half a century, many studies have been made on the Langevin model
which is widely employed as a useful model for a wide range
of stochastic phenomena (for a recent review, see Ref. \cite{Lindner04}).
Dynamics of a Brownian particle subjected to potential $V(x)$ 
is modeled by the Langevin equation given by
\begin{eqnarray}
\ddot{x} &=& - V'(x)-\gamma_0 \:\dot{x}+ \sqrt{2 D} \:\xi(t),
\label{eq:L1}
\end{eqnarray}
where dot and prime stand for derivatives with respect
to time and argument, respectively,
$\gamma_0$ denotes dissipation, $\xi(t)$ is zero-mean Gaussian white noise
with $\langle \xi(t) \xi(t') \rangle = \delta(t-t')$, and
$D$ expresses the strength of noise. 
Dissipation and diffusion terms satisfy the 
fluctuation-dissipation relation (FDR),
\begin{eqnarray}
D &=& k_B T \gamma_0,
\label{eq:L2}
\end{eqnarray}
where $k_B$ is the Boltzmann constant and $T$ the temperature.
The FDR implies that dissipation and diffusion processes 
originate from the same event.
The simple Langevin equation given by Eq. (\ref{eq:L1}) is based
on the two assumptions: (a) a dissipation is local in time and
(b) a diffusion depends on velocity $\dot{x}(t)$ but is independent of state $x(t)$. 
State-independent and state-dependent diffusions are commonly
referred to as additive and multiplicative noises, respectively. 
Multiplicative noise can be {\it phenomenologically} described in a number
of ways: for example, a diffusion term in Eq. (\ref{eq:L1}) may be generalized as
(for a review of study on multiplicative noise, see Ref.\cite{Munoz04})
\begin{eqnarray}
\sqrt{2 D} \xi(t) &\rightarrow& \sqrt{2D}\:G(x) \xi(t), 
\label{eq:L3}\\
&\rightarrow& \sqrt{2A} \xi_1(t)+ \sqrt{2M} G(x) \xi_2(t),
\label{eq:L4}
\end{eqnarray}
where $G(x)$ is a function of $x$, $A$ and $M$ denote strengths of
additive and multiplicative noises, respectively, and $\xi_1(t)$ and
$\xi_2(t)$ are white noises.
However, phenomenological diffusion terms given 
by Eqs. (\ref{eq:L3}) and (\ref{eq:L4}) 
have no microscopic bases and the FDR for such diffusions is not definite
\cite{Sakaguchi01}.
The stationary probability distribution function (PDF), 
which is obtained from the Fokker-Planck equation (FPE) 
corresponding to the Langevin model including these diffusion terms, 
is generally different from the Boltzmann factor, $e^{- V(x)/k_B T}$
\cite{Sakaguchi01,Sancho82,Anteneodo03,Hasegawa07}.

The importance of going beyond the assumptions (a) and (b) has been recognized
over many decades. 
The microscopic origin of additive noise has been proposed
within the framework of system-bath Hamiltonians
\cite{Ford65,Ullersma66,Caldeira81}, which are known as
Caldeira-Leggett (CL) type models. 
By using the generalized CL model including
a nonlinear system-bath coupling, we may obtain the non-Markovian
Langevin equation with nonlocal dissipation and multiplicative noise
which preserves the FDR \cite{Lindenberg81,Pollak93}.
The nature of nonlinear dissipation and multiplicative noise
has been recently explored with renewed interest 
\cite{Barik05,Chaudhuri06,Plyukhin07,Farias09}.
Ref. \cite{Barik05} discusses a possibility of observing a quantum current
in a system with quantum state-dependent diffusion and multiplicative noise.
Dynamics in a metastable state which is nonlinearly coupled to bath driven by
external noise has been studied \cite{Chaudhuri06}.
Ref. \cite{Plyukhin07} investigates a temporal development in an average velocity 
of noninteracting Brownian particles in a finite system with nonlinear dissipative force. 
Quite recently, 
a detailed comparison is made between non-Markovian and Markovian
Langevin equations including additive and multiplicative noises \cite{Farias09}.
It has been shown that in many cases, the Markovian (local) approximation
is not a reliable description of the non-Markovian (non-local) dynamics
\cite{Farias09}. Nonlinear dissipation and multiplicative noise
have been recognized as important ingredients in several fields
such as mesoscopic scale systems \cite{Zaitsev09,Eichler11} 
and ratchet problems \cite{Magnasco93,Julicher97,Reimann02,Porto00}.

In the last decade, we have significant progress in experimental
and theoretical understanding of nonequilibrium statistics 
(for reviews, see Refs. \cite{Busta05,Ritort07,Ciliberto10}).
At the moment we have three kinds of fluctuation theorems:
the Jarzynski equality (JE) \cite{Jarzynski97}, 
the steady- and transient-fluctuation theorem \cite{Evans93,Narayan04,Crooks99}, 
and the Crooks theorem \cite{Narayan04,Crooks99}. 
These theorems are applicable to nonequilibrium systems driven arbitrarily 
far from the equilibrium state.
In this paper, we pay our attention to the JE which was originally
proposed for a classical isolated system and open system weakly coupled
to baths \cite{Jarzynski97,Jarzynski97b}. 
Subsequently Jarzynski proved that the JE 
is valid for strongly coupled classical open systems \cite{Jarzynski04}. 
A validity of the JE has been confirmed by various experiments 
for systems which may be described by damped harmonic oscillator models
\cite{Liphardt02,Wang05,Douarche05,Douarche06,Joubaud07,Joubaud07b}.
Stimulated by these experiments, many theoretical analyses have been made for 
harmonic oscillators with the use of the Markovian Langevin model 
\cite{Douarche05,Douarche06,Joubaud07,Joubaud07b},
the non-Markovian Langevin model \cite{Zamponi05,Mai07,Speck07,Ohkuma07},
Fokker-Planck equation \cite{Chaudhury08}, and Hamiltonian model 
\cite{Jarzynski06,Jarzynski08,Dhar05,Chakrabarti08,Hijar10,Hasegawa11b}.
The validity of the JE has been examined for anharmonic oscillators 
\cite{Mai07,Saha06} and for van der Pol and Rayleigh oscillators \cite{Hasegawa11c}.
We should note that these studies have been made for the non-Markovian and/or
Markovian Langevin models with additive noise. Recently the JE in the Markovian 
Langevin model with multiplicative white noise for Brownian particles has been discussed
in Ref. \cite{Lev10}. However, a study on the JE for the non-Markovian Langevin model 
with multiplicative {\it colored} noise is scanty at the moment \cite{Aron10}.

The purpose of the present paper is twofold: (1) to make a detailed study of 
the non-Markovian Langevin model derived from the generalized CL model for classical open systems 
with nonlinear nonlocal dissipation and state-dependent diffusion and
(2) to calculate responses to applied forces and examine a validity of the JE in the system. 
In the following Sec. II, we derive the non-Markovian Langevin equation,
adopting the generalized CL model including nonlinear system-bath coupling
\cite{Caldeira81,Lindenberg81,Pollak93,Barik05} (Sec. IIA).
The Ornstein-Uhlenbeck (OU) process of colored noise is taken into account.
By using the two methods \cite{Pollak93,Farias09,Bao05} in which
new variables are introduced into the non-Markovian Langevin equation,
we obtain a set of four first-order differential equations 
and the relevant multi-variate FPE. 
The local limit of the non-Markovian Langevin equation is examined (Sec. IIB).
In Sec. III, we study harmonic oscillator systems, 
applying simulation method to the four differential equations
mentioned above. We calculate the stationary marginal PDF of the system (Sec. IIIA) 
and its responses to applied sinusoidal and step forces (Sec. IIIB). 
In particular, frequency-dependent responses of the mean position 
of $\langle x(t) \rangle$ to sinusoidal force have been made in detail.
We obtain enhanced fluctuations of $\langle [x(t)-\langle x(t) \rangle]^2 \rangle$ 
induced by a combined effect of nonlinear dissipation and multiplicative noise. 
Applying a ramp force to the system, the validity of the JE has been examined (Sec. IIIC).
In Sec. IV we discuss the over-damped limit of the Markovian Langevin model
subjected to multiplicative noise.
Sec. V is devoted to our conclusion.

\section{The system-bath model}
\subsection{Non-Markovian Langevin equation}

We consider a system of a classical oscillator coupled to 
a bath consisting of $N$-body uncoupled oscillators described by
the CL model \cite{Caldeira81,Barik05},
\begin{eqnarray}
H &=& H_S + H_B + H_I,
\label{eq:A1}
\end{eqnarray}
with
\begin{eqnarray}
H_S &=& \frac{p^2}{2}+ V(x) - x f(t), 
\label{eq:A1b} \\
H_B+H_I &=& \sum_{n=1}^N \left\{ \frac{p_n^2}{2 m_n}
+ \frac{m_n \omega_n^2}{2}\left( q_n -\frac{c_n \phi(x)}{m_n \omega_n^2}\right)^2 \right\}.
\label{eq:A1c} 
\end{eqnarray}
Here $H_S$, $H_B$ and $H_I$ express Hamiltonians of the system, bath and interaction, 
respectively: $x$, $p$ and $V(x)$ denote position, momentum and potential, respectively, 
of the system: $q_n$, $p_n$, $m_n$ and $\omega_n$ stand for position, momentum, mass 
and frequency, respectively, of bath: the system couples to the bath nonlinearly 
through a function $\phi(x)$: 
$f(t)$ expresses an applied external force.
The original CL model adopts a linear system-bath coupling
with $\phi(x)=x$ in Eq. (\ref{eq:A1c}) which yields additive noise \cite{Caldeira81}.
By using the standard procedure, we obtain the generalized Langevin 
equation given by \cite{Caldeira81,Lindenberg81,Pollak93,Barik05}
\begin{eqnarray}
\ddot{x}(t) &=& -V'(x(t))
-\phi'(x(t)) \int_0^t \gamma(t-t') \:\phi'(x(t')) \:\dot{x}(t')\:dt'
+\phi'(x(t))\:\zeta(t)+f(t),
\label{eq:A4}
\end{eqnarray}
with
\begin{eqnarray}
\gamma(t) &=& \sum_{n=1}^N \left(\frac{c_n^2}{m_n \omega_n^2} \right) \cos \omega_n t, 
\label{eq:A5}\\
\zeta(t) 
&=& \sum_{n=1}^N \left\{ \left[\frac{m_n \omega_n^2}{c_n}\: q_n(0) 
-\phi(x(0)) \right]\left( \frac{c_n^2}{m_n\omega_n^2} \right) \cos \omega_n t
+\left( \frac{c_n p_n(0)}{m_n \omega_n} \right) \sin \omega_n t \right\}, 
\label{eq:A6}
\end{eqnarray}
where $\gamma(t-t')$ denotes the non-local kernel and $\zeta(t)$ stands for noise.
Dissipation and diffusion terms given by Eqs. (\ref{eq:A5}) and (\ref{eq:A6}),
respectively, satisfy the second-kind FDR,
\begin{eqnarray}
\left< \zeta(t) \zeta(t') \right>_0
&=& k_B T \: \gamma(t-t'),
\label{eq:A7}
\end{eqnarray}
where the bracket $\langle \cdot \rangle_0$ stands for the average over initial states 
of $q_n(0)$ and $p_n(0)$ \cite{Lindenberg81,Pollak93,Barik05}.

We have adopted the OU process for the kernel $\gamma(t-t')$ given by
\begin{eqnarray}
\gamma(t-t') &=& \left( \frac{\gamma_0}{\tau} \right)
e^{-(t-t')/\tau},
\label{eq:A8}
\end{eqnarray}
where $\tau$ and $\gamma_0$ stand for the relaxation time and strength,
respectively, of colored noise.
The OU colored noise may be generated by the differential equation,
\begin{eqnarray}
\dot{\zeta}(t) &=& -\frac{\zeta(t)}{\tau}
+\frac{\sqrt{2 k_B T \gamma_0} }{\tau} \: \xi(t),
\label{eq:A14b}
\end{eqnarray}
where $\xi(t)$ expresses white noise with
\begin{eqnarray}
\left< \xi(t) \right> &=& 0, \;\;\;\;
\left< \xi(t) \xi(t')\right> = \delta(t-t').
\label{eq:A15}
\end{eqnarray}
Equations (\ref{eq:A14b}) and (\ref{eq:A15}) lead to the PDF
and correlation of colored noise given by
\begin{eqnarray}
P(\zeta) &\propto& e^{-(\beta \tau/2 \gamma_0) \:\zeta^2},
\label{eq:A17} \\
\left< \zeta(t) \zeta(t')\right> 
&=& \left( \frac{k_B T \gamma_0}{\tau}\right) \:e^{-(t-t')/\tau}
= k_B T \:\gamma(t-t'), 
\label{eq:A16}
\end{eqnarray}
where $\beta=1/k_B T$.

\subsubsection{The method A}
The two methods have been proposed to transform the non-Markovian Langevin
equation given by Eq. (\ref{eq:A4}) into multiple differential equations \cite{Pollak93,Farias09,Bao05}.
In the method A, we introduce a new variable $u(t)$ \cite{Farias09,Bao05},
\begin{eqnarray}
u(t) &=& - \int_0^t \gamma(t-t') \phi'(x(t')) \dot{x}(t')\:dt',
\label{eq:A10}
\end{eqnarray}
to obtain four first-order differential equations for $x(t)$, $p(t)$, $u(t)$ 
and $\zeta(t)$ given by
\begin{eqnarray}
\dot{x}(t) &=& p(t), 
\label{eq:A11}\\
\dot{p}(t) &=& -V'(x)+\phi'(x(t)) \:u(t) +f(t)
+ \phi'(x(t)) \:\zeta(t), 
\label{eq:A12}\\
\dot{u}(t) &=& - \frac{u(t)}{\tau} 
- \left( \frac{\gamma_0}{\tau} \right) \phi'(x(t)) \:p(t), 
\label{eq:A13}\\
\dot{\zeta}(t) &=& -\frac{\zeta(t)}{\tau}
+\frac{\sqrt{2 k_B T \gamma_0} }{\tau} \: \xi(t).
\label{eq:A14}
\end{eqnarray}

From Eqs. (\ref{eq:A11})-(\ref{eq:A14}), we obtain the multi-variate FPE 
for distribution of $P(x,p,u,\zeta,t)$,
\begin{eqnarray}
\frac{\partial P(x,p,u,\zeta,t)}{\partial t} 
&=& - \frac{\partial}{\partial x} \:p \: P(x,p,u,\zeta,t) \nonumber \\
&-& \frac{\partial}{\partial p}\left[-V'(x)+f(t)+\phi'(x) u+\phi'(x)\zeta
\right]P(x,p,u,\zeta,t) \nonumber \\
&-& \frac{\partial}{\partial u}\left[ \frac{u}{\tau}
+\left( \frac{\gamma_0}{\tau} \right) \:\phi'(x) p \right] P(x,p,u,\zeta,t)
+ \frac{\partial}{\partial \zeta} \left( \frac{\zeta}{\tau} \right)
P(x,p,u,\zeta,t) \nonumber \\
&+& \left( \frac{k_B T \gamma_0}{\tau^2} \right) \frac{\partial^2}{\partial \zeta^2} 
P(x,p,u,\zeta,t). 
\label{eq:A18}
\end{eqnarray}

\subsubsection{The method B}

In the method B, we introduce a new variable $z(t)$ \cite{Pollak93},
\begin{eqnarray}
z(t) &=& - \left( \frac{\tau}{\gamma_0} \right)
\int_{0}^{t} \gamma(t-t')\phi'(x(t'))\:dt' 
+\left( \frac{\tau}{\gamma_0}\right) \zeta(t)+\phi(t),
\label{eq:Y1}
\end{eqnarray}
to obtain four first-order differential equations,
\begin{eqnarray}
\dot{x}(t) &=& p(t),
\label{eq:Y2}\\
\dot{p}(t) &=& -\frac{\partial U(x, z)}{\partial x}+f(t),
\label{eq:Y3} \\
\dot{z}(t) &=& -\frac{1}{\gamma_0} \frac{\partial U(x, z)}{\partial z}
+ \frac{\sqrt{ 2 k_B T \gamma_0}}{\gamma_0} \:\xi(t), 
\label{eq:Y4}\\
%
\dot{\zeta}(t) &=& - \frac{\zeta(t)}{\tau} 
+ \frac{\sqrt{2 k_B T \gamma_0}}{\tau} \:\xi(t),
\label{eq:Y5}
\end{eqnarray}
with
\begin{eqnarray}
U(x, z) &=& V(x)+\left( \frac{\gamma_0}{2 \tau} \right) [z-\phi(x)]^2.
\label{eq:Y6}
\end{eqnarray}
It is noted that white noises in Eqs. (\ref{eq:Y4}) and (\ref{eq:Y5})
come from the same origin.

A variable $\zeta(t)$ in Eq. (\ref{eq:Y5}) is isolated from the rest of variables
in the four differential equations.
From Eq. (\ref{eq:Y5}) we may obtain its stationary PDF, $P(\zeta)$, 
given by Eq. (\ref{eq:A17}).
The FPE relevant to Eqs. (\ref{eq:Y2})-(\ref{eq:Y4}) for $P(x, p, z, t)$ 
is given by
\begin{eqnarray}
\frac{\partial P(x, p, z)}{\partial t} 
&=& - \frac{\partial}{\partial x} p  P(x, p, z)
+ \frac{\partial }{\partial p} 
\left[\frac{\partial U(x,z)}{\partial x} -f(t) \right] P(x, p, z)
+ \frac{1}{\gamma_0}\frac{\partial}{\partial z} 
\frac{\partial U(x,z)}{\partial z}P(x,p,z)
\nonumber \\
&+&\frac{k _B T}{\gamma_0} \frac{\partial^2}{\partial z^2}P(x,p,z).  
\label{eq:Y7} 
\end{eqnarray}
The stationary PDF of Eq. (\ref{eq:Y7}) with $f(t)=0$ is given by \cite{Pollak93}
\begin{eqnarray}
P(x, p, z) &\propto& e^{-\beta [p^2/2+U(x,z)]},
\label{eq:Y8}
\end{eqnarray}
which leads to stationary marginal PDFs,
\begin{eqnarray}
P(x) &=& \int P(x,p,z) \:dp \:dz 
\propto e^{-\beta V(x)}, 
\label{eq:A19} \\
P(p) &=& \int P(x,p,z) \:dx \:dz
\propto e^{-\beta p^2/2}.
\label{eq:A20}
\end{eqnarray}

Equations (\ref{eq:A11})-(\ref{eq:A14}) in the method A are equivalent 
to Eqs. (\ref{eq:Y2})-(\ref{eq:Y5}) in the method B.
An advantage of the method A is that the local limit of $\tau \rightarrow 0$ 
is easily obtainable in Eqs. (\ref{eq:A11})-(\ref{eq:A14}), 
while in the method B the analytical expression for the stationary PDF 
given by Eq. (\ref{eq:Y8}) may be derived.
In our simulations to be reported in the following section,
we have mainly employed the method A, whose results are
partly checked by separate simulations using the method B.

\subsection{Markovian Langevin equation}
It is worthwhile to examine the local limit of Eq. (\ref{eq:A4})
with a kernel $\gamma(t-t')$ given by
\begin{eqnarray}
\gamma(t-t') &=&  2 \gamma_0 \: \delta(t-t'),
\label{eq:B1}
\end{eqnarray}
which leads to the Markovian Langevin equation,
\begin{eqnarray}
\ddot{x}(t) &=& -V'(x(t)) -\gamma_0 \phi'(x(t))^2 \:\dot{x}(t)
+\sqrt{2 k_B T \gamma_0} \:\phi'(x(t))  \:\xi(t) +f(t).
\label{eq:B2}
\end{eqnarray}
It is evident that the Markovian Langevin equation becomes a good approximation
of the non-Markovian one in the limit of $\tau \rightarrow 0$.

From Eq. (\ref{eq:B2}), we obtain three differential equations,
\begin{eqnarray}
\dot{x}(t) &=& p(t), \\
\dot{p}(t) &=& -V'(x)-\gamma_0 \phi'(x(t))^2 p(t) +f(t)
+ \phi'(x(t)) \:\zeta(t), \\
\dot{\zeta}(t) &=& -\frac{\zeta(t)}{\tau}
+\frac{\sqrt{2 k_B T \gamma_0} }{\tau} \: \xi(t).
\end{eqnarray}
The PDF for $\zeta$ is given by Eq. (\ref{eq:A17}).
The relevant FPE for the PDF of $P(x,p,t)$ is expressed by
\begin{eqnarray}
\frac{\partial P(x,p,t)}{\partial t} 
&=& - \frac{\partial }{\partial x} \:p \:P(x,p,t) 
+ \frac{\partial}{\partial p}
\left[ V'(x)-f(t)+\gamma_0 \phi'(x)^2 \:p \right] P(x,p,t)
\nonumber  \\
&+& k_B T \:\gamma_0 \:\phi'(x)^2 
\frac{\partial}{\partial p} \phi'(x) 
\frac{\partial}{\partial p} \phi'(x) P(x,p,t).
\label{eq:B3}
\end{eqnarray}
The stationary distribution of Eq. (\ref{eq:B3}) with $f(t)=0$ is given by
\begin{eqnarray}
P(x, p) &\propto &  \:e^{-\beta[p^2/2+V(x)]}.
\label{eq:B4}
\end{eqnarray}
This is consistent with the result of the non-Markovian Langevin
equation given by Eq. (\ref{eq:Y8}).

\section{Harmonic oscillator systems}
\subsection{Stationary marginal PDF}

Simulations have been performed  for harmonic oscillator systems 
where $V(x)$ and $\phi(x)$ in Eqs. (\ref{eq:A1})-(\ref{eq:A1c}) are given by
\begin{eqnarray}
V(x) &=& \frac{\omega_s^2 x^2}{2}, \\
\phi(x) &=& \frac{a x^2}{2} + b x,
\end{eqnarray}
where $\omega_s$ stands for oscillator frequency of the system, and
$a$ and $b$ denote magnitudes of multiplicative and additive noises, respectively.
We have solved Eqs. (\ref{eq:A11})-(\ref{eq:A14}) by using
the Heun method \cite{Note1} with a time step of 0.001 for parameters of 
$\omega_s=1.0$, $\gamma_0=1.0$ and $k_B T=1.0$ otherwise noticed.
Simulations have been made for $0 \leq t < 1000.0$,
which are averaged over $10^4$ sets of initial states of
Gaussian-distributed $x(0)$ and $p(0)$
with $\langle x(0) \rangle= \langle p(0) \rangle=0$ and 
$\langle p(0)^2 \rangle = \langle \omega_s^2 x(0)^2 \rangle = k_B T$. 
In all simulations, we have used the initial conditions of $u(0)=0$ and $\zeta(0)=0$.
 
First we show marginal PDFs of $P(x)$, $P(p)$, $P(u)$ and $P(\zeta)$ 
for $f(t)=0$, which are evaluated by simulations of Eqs. (\ref{eq:A11})-(\ref{eq:A14}) 
with discarding initial results at $t < 200.0$.
Figures \ref{fig1}(a), (b), (c) and (d) show $P(x)$, $P(p)$, $P(u)$
and $P(\zeta)$, respectively, obtained 
for $(a,b,\tau)=(1, 0,10.0)$ (solid curves), $(1,0,0.1)$ (dashed curves),
$(0,1,10.0)$ (chain curves) and $(0,1,0.1)$ (dotted curves),
where 
$\tau$ stands for the relaxation time of OU colored noise.
We note that $P(x)$ and $P(p)$ in Figs. \ref{fig1}(a) and (b) are independent of
the parameters of $(a, b, \tau)$ although $P(u)$ and $P(\zeta)$ 
in Figs. \ref{fig1}(c) and (d) depend on them.
Calculated $P(x)$ and $P(p)$ are in good agreement with
Gaussian PDFs given by Eqs. (\ref{eq:A19}) and (\ref{eq:A20}).
Equation (\ref{eq:A17}) shows that $P(\zeta)$ is the Gaussian PDF whose variance
depends on $\tau$ for fixed $\gamma_0$ and $T$.
In contrast, Fig. \ref{fig1}(c) shows that $P(u)$ is Gaussian PDF
for additive noise  but non-Gaussian PDF for multiplicative noise:
the kurtosis of $P(u)$ defined by
\begin{eqnarray}
\eta_u &=& \frac{\left< (u-\left< u \right>_u)^4\right>_u}
{\left< (u-\left< u \right>_u)^2\right>_u^2},
\label{eq:Y9}
\end{eqnarray}
is $\eta_u=3.0$, 3.0, 7.8 and 10.3 for $(a,b, \tau)=(0,1,0.1)$, $(0,1,10.0)$, 
$(1, 0, 0.1)$ and $(1, 0, 10.0)$, respectively, where the bracket 
$\langle \cdot \rangle_u$ denotes an average over $P(u)$.
Note that $\eta_u=3.0$ for the Gaussian distribution.

\subsection{Responses to applied forces}

\subsubsection{Sinusoidal forces}
Dynamical responses of harmonic oscillator systems to applied sinusoidal 
and step forces are studied. We first apply a sinusoidal force given by
\begin{eqnarray}
f(t) &=& g \:\sin\left(\frac{2 \pi t}{T_0} \right) =g\:\sin\omega_0 t,
\label{eq:C1}
\end{eqnarray}
where $g$, $T_0$ and $\omega_0$ ($=2 \pi/T_0$) denote 
its magnitude, period and frequency, respectively.
A sinusoidal force given by Eq. (\ref{eq:C1}) yields 
averaged outputs given by
\begin{eqnarray}
\mu_x(t) &=& \langle x(t) \rangle, \\
\mu_p(t) &=& \langle p(t) \rangle = \dot{\mu}_x(t),
\label{eq:C2}
\end{eqnarray}
Fourier-transformed outputs given by
\begin{eqnarray}
\mu_x[\omega] &=& \int_{-\infty}^{\infty} e^{i\:\omega t} \mu_x(t) \:dt, \\
\mu_p[\omega] &=& \int_{-\infty}^{\infty} e^{i\:\omega t} \mu_p(t) \:dt,
\label{eq:C3}
\end{eqnarray}
have peaks at $\omega=\omega_0$.
Output magnitudes defined by
\begin{eqnarray}
I_x(\omega_0) &=& \vert \mu_x[\omega_0] \vert^2, 
\label{eq:C4} \\
I_p(\omega_0) &=& \vert \mu_p[\omega_0] \vert^2
= \omega_0^2 \:I_x(\omega_0),
\label{eq:C4b}
\end{eqnarray}
express frequency-dependent responses of the system.

In the case of additive noise only [$\phi(x)=x$], 
we may obtain the analytical result of $I_x(\omega_0)$
with the susceptibility given by
\begin{eqnarray}
R_x[\omega] &=& \frac{\mu_x[\omega]}{f[\omega]}
= \frac{1}{\omega_s^2 -\omega^2 - i \:\omega \gamma[\omega]},
\label{eq:C5}
\end{eqnarray}
where $\gamma[\omega]$ is the Fourier transform of 
the kernel, $\gamma(t)$,
\begin{eqnarray}
\gamma[\omega] &=& 
\frac{\gamma_0}{1-i\:\omega \tau}.
\label{eq:C6} 
\end{eqnarray}
From Eqs. (\ref{eq:C4})-(\ref{eq:C6}), the output magnitude is given by
\begin{eqnarray}
I_x(\omega_0) &=& \frac{\pi^2 g^2}
{(\omega_0^2-\omega_s^2)^2+\omega_0^2 \vert \gamma[\omega_0] \vert^2},\\
&=& \frac{\pi^2 g^2 (1+\omega_0^2 \tau^2)}
{[(\omega_0^2-\omega_s^2)^2(1+\omega_0^2 \tau^2)+ \omega_0^2 \gamma_0^2]}
\hspace{1cm}\mbox{for $\omega_0 > 0$},
\label{eq:C7} \\
&= & \frac{\pi^2 g^2(1+\omega_0^2 \tau^2)}{\omega_0^2 \gamma_0^2}
\hspace{3cm}\mbox{for $\omega_0 = \omega_s$}.
\label{eq:C7b}
\end{eqnarray}
Equation (\ref{eq:C7b}) shows that $I_x(\omega_0)$ at a resonance frequency 
($\omega_0=\omega_s$) is monotonously increased with increasing $\tau$.

In the case of a general $\phi(x)$ yielding multiplicative noise, however, 
we cannot make an analytical study because the Fourier or Laplace transformation
cannot be employed. Then we have to rely on numerical simulations
of Eqs. (\ref{eq:A11})-(\ref{eq:A14}).
Figures \ref{fig2}(a) and (b) show time courses of $\mu_x(t)$
for $\tau=0.1$ and $\tau=10.0$, respectively, 
with $(a, b)=(0, 1)$ and $(1, 0)$ when a sinusoidal force 
given by Eq. (\ref{eq:C1}) with $g=1.0$ and $T_0=10.0$ is applied.
The results of $(a,b)=(1,0)$ and $(0,1)$ are almost the same for $\tau=0.1$ in Fig. 2(a).
In contrast, for $\tau=10.0$, $\mu_x(t)$ of $(a,b)=(1,0)$ is a little different
from that of $(a,b)=(0,1)$ in Fig. 2(b): an irregularity in the former is larger than 
that of the latter. The irregularity in $\mu_x(t)$ for $(a,b)=(0,1)$
is gradually reduced at a larger $t$ (relevant result not shown).

Although the difference between $\mu_x(t)$ for different $\tau$ values
with additive and multiplicative noises is not so clear in Fig. 2,
it becomes evident in the Fourier-transformed quantity of 
$\mu_x[\omega]$ or $I_x(\omega_0)$.
Figure \ref{fig3}(a) shows the $\omega_0$ dependence of output magnitudes of $I_x(\omega_0)$
with $(a, b)=(0, 1)$ (chain curve) and $(1, 0)$ (solid curve) for $\tau=10.0$. 
The chain curve for additive noise with $(a, b)=(0, 1)$
has a resonance peak at the frequency of a bath oscillator ($\omega=\omega_s=1.0$).
The dashed curve expresses a theoretical result
for additive noise calculated by Eq. (\ref{eq:C7}), which is in good agreement
with a relevant result of simulations except for $\omega_0 \sim \omega_s$
where our simulation overestimates $I_x(\omega_0)$.
In contrast, the solid curve in Fig. \ref{fig3}(a) for multiplicative noise 
has a $\lambda$-type peak at $\omega_0 = \omega_r \sim 1.4$ 
which is different from $\omega_s$.
Furthermore, a shape of the solid curve is rather peculiar and different
from that of the chain curve for additive noise.
Figure \ref{fig3}(b) shows a similar plot of $I_x(\omega_0)$ for $\tau=0.1$
with $(a, b)=(0, 1)$ (chain curve) 
and $(1, 0)$ (solid curve). 
A broad peak at $\omega_0 \sim 1$ for multiplicative noise is smaller 
than that for additive noise.

Figures \ref{fig4}(a)-(d) show the Lissajous plots 
of $\mu_x(t)$ versus $f(t)$ with multiplicative noise of $(a, b, \tau)=(1, 0, 10.0)$
for $T_0=10.0$, 5.0, 4.0 and 3.0 which correspond to
$\omega_0=0.63$, 1.26, 1.57 and 2.09, respectively.
We note that $\mu_x(t)$ is in phase with $f(t)$ for $T_0=10.0$
but in anti-phase for $T_0=3.0$, and that the transition
from an in-phase to an anti-phase occurs at $T_0 \sim 4.5$
($\omega_0 = 1.4 = \omega_r$). This is consistent with a peak
position of $I_x(\omega_0)$ for multiplicative noise shown in Fig. \ref{fig3}(a).

We study fluctuations of $x(t)$ in the system defined by
\begin{eqnarray}
\rho_x(t) &=& \langle [x(t)-\langle x(t) \rangle]^2 \rangle.
\end{eqnarray}
It is noted that in stationary state without forces, 
we obtain $\rho_x(t)=\rho_p(t)=k_B T$
in harmonic oscillator systems 
because marginal stationary PDFs
are given by $P(x) \propto e^{-\beta x^2/2}$ and $P(p)\propto e^{-\beta p^2/2}$ 
regardless of values of $a$, $b$ and $\tau$.
It is not the case when sinusoidal forces are applied to the system 
with multiplicative noise, as will be shown in the following.
Figures \ref{fig5}(a) and (b) show time courses of
$\rho_x(t)$ in harmonic oscillator systems 
for $\tau=0.1$ and $\tau=10.0$, respectively,
for multiplicative noise of $(a, b)=(1, 0)$ when sinusoidal forces 
with $g=1.0$ and $T_0=6.0$, 10.0 and 100.0 are applied.
All $\rho_s(t)$ start from $\rho_s(0)=1.0$ at $t=0.0$, results for $T_0=6.0$ and 10.0
being shifted by four and two, respectively, for a clarity of the figure.
For $T_0=100.0$, we obtain $\rho_x(t) \simeq 1.0$ for $\tau=0.1$ and 10.0. 
In contrast, $\rho_x(t)$ for $T_0=10.0$ and $6.0$ is much increased than unity.
This is more clearly seen in Figs. \ref{fig6}(a) and (b), where
the stationary value of $\rho_{xs}$ defined by
\begin{eqnarray}
\rho_{xs} &\equiv& \rho_x(t)
\hspace{1cm}\mbox{at $t \sim 1000.0$},
\end{eqnarray}
is plotted as a function of $\omega_0$ for additive (dashed curves) and
multiplicative noise (solid curves).
The solid curve in Fig. \ref{fig6}(a) expressing $\rho_{xs}$ 
for $\tau=10.0$ with multiplicative noise has a peak at $\omega_0 \sim \omega_r$
where $\omega_r \simeq 1.4$ discussed before.
The magnitude of a peak for $\tau=0.1$ with multiplicative noise 
in Fig. \ref{fig6}(b) is less significant than that for $\tau=10.0$ 
in Fig. \ref{fig6}(a).
On the other hand, dashed curves in Figs. \ref{fig6}(a) and \ref{fig6}(b) expressing
$\rho_x(t)$ for additive noise of $(a, b)=(0,1)$ are given by 
$\rho_x(t)=k_B T=1.0$ independently of $T_0$ and $\tau$.

Solid and chain curves in Fig. \ref{fig7} express the $\tau$ dependence of $\rho_{xs}$ 
with $T_0=5.0$ and  $T_0=10.0$, respectively, for multiplicative noise.
With increasing $\tau$, $\rho_{xs}$ with $T_0=5.0$ is much increased
than that with $T_0=10.0$ because the former is closer to 
$2 \pi/\omega_r \sim 4.19$ than the latter. 
For additive noise, we obtain $\rho_{xs}=1.0$ (dashed curve)
as mentioned above.
These enhanced fluctuations arise from a combined effect of nonlinear
dissipation and multiplicative colored noise. 

\subsubsection{Step forces}
Next we apply a step force given by
\begin{eqnarray}
f(t) &=& \left\{ \begin{array}{ll}
0
\quad & \mbox{for $t < t_1 $}, \\ 
g
\quad & \mbox{for $t \geq t_1 $},
\end{array} \right. 
\label{eq:C8}
\end{eqnarray}
where $t_1$ is the starting time of an force with a magnitude of $g$.
Figure \ref{fig8}(a) shows time courses
of $\mu_x(t)$ 
with additive noise with $(a, b)=(0, 1)$ for $\tau=0.1$, 1.0 and 10.0
when step forces with $g=1.0$ are applied at $t = t_1=100.0$.
For $\tau=10.0$, an oscillation induced by a step force applied at
$t=100.0$ remains for a fairly long period after $t_1$, whereas
those for $\tau=1.0$ and $0.1$ are quickly damped out.
Figure \ref{fig8}(b) shows similar time courses of $\mu_x(t)$
for multiplicative noise with $(a, b)=(1, 0)$ for $\tau=0.1$, 1.0 and 10.0.
Comparing Fig. \ref{fig8}(a) with Fig. \ref{fig8}(b), we notice that
an oscillation with multiplicative noise is more quickly
disappear than that with additive noise.

Figure \ref{fig9} shows $\mu_x(t)$ for four sets of $(a, b)=(0.0, 1.0)$,
(0.2, 0.8), (0.5, 0.5) and (1.0, 0.0) with $\tau=10.0$. 
With increasing a component of the multiplicative noise, the oscillation induced 
by an applied step force is rapidly decayed.

\subsection{The Jarzynski equality}

The JE is expressed by \cite{Jarzynski97,Jarzynski04}
\begin{eqnarray}
e^{-\beta \Delta F} &=& \langle e^{-\beta W} \rangle_W
= \int e^{-\beta W}\: P(W)  \:dW,
\label{eq:J1}
\end{eqnarray}
where $W$ stands for a work made in a system when its parameter is changed,
the bracket $\langle \cdot \rangle_W$ means the average over the
work distribution function (WDF), $P(W)$, of a work performed by a prescribed protocol,
and $\Delta F$ denotes the free-energy difference between the initial and
final equilibrium states [Eqs. (\ref{eq:J4}) and (\ref{eq:J5})].
Equation (\ref{eq:J1}) includes the second law of the thermodynamics,
$\langle W \rangle_W \geq \Delta F$, where the equality holds for the
reversible process.
The JE in Eq. (\ref{eq:J1}) may be rewritten as
\begin{eqnarray}
R &\equiv& - \frac{1}{\beta} \ln \langle e^{-\beta W} \rangle_W = \Delta F.
\label{eq:J2}
\end{eqnarray} 
When the WDF is Gaussian given by
\begin{eqnarray}
P(W) &=& \frac{1}{\sqrt{2 \pi \sigma_W^2}}
\:e^{-(W-\mu_W)^2/2 \sigma_W^2},
\label{eq:J3}
\end{eqnarray}
we obtain 
\begin{eqnarray}
R &=& \mu_W-\frac{\beta \sigma_W^2}{2},
\label{eq:J3b}
\end{eqnarray}
with
\begin{eqnarray}
\mu_W &=& \left< W \right>_W, \\
\sigma_W^2 &=& \left< (W-\mu_W)^2 \right>_W,
\end{eqnarray}
where $\mu_W$ and $\sigma_W^2$ express mean and variance, respectively, of the WDF. 
Equation (\ref{eq:J3b}) is not valid when the WDF is non-Gaussian.

We apply a ramp force given by
\begin{eqnarray}
f(t) &=& \left\{ \begin{array}{ll}
0
\quad & \mbox{for $t < 0 $}, \\ 
g (\frac{t}{\tau_f})
\quad & \mbox{for $0 \leq t < \tau_f $}, \\ 
g
\quad & \mbox{for $t \geq \tau_f $},
\end{array} \right. 
\label{eq:J9}
\end{eqnarray}
where $\tau_f$ stands for a duration of the applied force with a magnitude $g$ ($=2.0$).

The free energy difference $\Delta F$ between equilibrium states with $f(t)=0$
and $f(t)=g$ ($g$, constant) is given by \cite{Jarzynski04,Hasegawa11c}
\begin{eqnarray}
\Delta F &=& F(g)-F(0) 
= - \frac{1}{\beta} \ln \left( \frac{Z_S(g)}{Z_S(0)} \right),
\label{eq:J4}
\end{eqnarray}
with 
\begin{eqnarray}
Z_S(g) &=& \frac{{\rm Tr}\:\{e^{-\beta[H_S(g)+H_B+H_I]} \}} 
{{\rm Tr}\:\{e^{-\beta H_B}\} }, 
\label{eq:J5}
\end{eqnarray}
where $H_S(g)$ denotes the system Hamiltonian with $f(t)=g$.
After some manipulations (detail being given in the Appendix), we obtain
\begin{eqnarray}
\Delta F &=& -\frac{g^2}{2 \omega_s^2},
\label{eq:J8}
\end{eqnarray}
independently of $a$, $b$ and $\tau$, which becomes $\Delta F=-2.0$ 
for $\omega_s=1.0$ and $g=2.0$.

Simulations have been performed with the same parameters as in Secs. III A and
III B, but over $10^5$ sets of initial states.
Simulation results are presented in Figs. \ref{fig10}-\ref{fig12}.
Figure \ref{fig10}(a), \ref{fig10}(b) and \ref{fig10}(c) show WDFs for ramp forces of
$\tau_f=100.0$, 10.0 and 1.0, respectively, 
with four sets of parameters of $(a, b, \tau)$ $=(1, 0, 10.0)$ (solid curves), 
(1, 0, 0.1) (dashed curves), (0, 1, 10.0) (chain curves),  and (0, 1, 0.1) (dotted curves).
We note in Fig. \ref{fig10}(a) that all WDFs for $\tau=100.0$ locate  
at $\mu_W \sim -2.0$ with widths of $\sigma_W \sim 0.3$ and that WDFs 
for multiplicative noise are asymmetric non-Gaussian, 
while those for additive noise are (symmetric) Gaussian. 
Indeed, the kurtosis of the WDF given by
\begin{eqnarray}
\eta_W &=& \frac{\left< (W-\mu_W)^4\right>_W}{(\sigma_W^2)^2},
\label{eq:Y10}
\end{eqnarray}
is $\eta_W=3.0$, 3.0, 3.9 and 4.2 for $(a, b, \tau)=(0, 1, 0.1)$, $(0, 1, 10.0)$,
$(1, 0, 0.1)$, and $(1, 0, 10.0)$, respectively, with $\tau_f=100.0$.
When $\tau_f$ is reduced to 10.0, behaviors of $P(W)$ are changed.
Figure \ref{fig10}(b) shows that WDFs for $\tau=10.0$ 
with different sets of $(a, b)$ show different behavior but 
with almost the same values of $\mu_W \sim -1.8$ and $\sigma_W \sim 1.0$.
WDFs for multiplicative noise much departs from Gaussian distribution:
the kurtosis of $P(W)$ for $\tau=10.0$ is $\eta_W=3.0$, 3.0, 7.0 and 5.6 
for $(a, b, \tau)=(0, 1, 0.1)$, $(0, 1, 10.0)$,
$(1, 0, 0.1)$, and $(1, 0, 10.0)$, respectively.
When $\tau_f$ is further reduced to 1.0, 
we note in Fig. \ref{fig10}(c) that
all WDFs become almost identical Gaussian ($\eta_W \simeq 3.0$) with
$\mu_W \sim 0.0$ and $\sigma_W \sim 2.0$.

These changes in $\mu_W$, $\sigma_W$ and $\eta_W$ as a function of $\tau_f$ are
shown in Figs. \ref{fig11}(a), \ref{fig11}(b) and \ref{fig11}(c), respectively,
where marks express simulation results and curves are plotted only
for a guide of the eye.
With decreasing $\tau$ from 100.0 to 0.1, $\mu_W$ changes from $\mu_W \sim -2.0$
to 0.0 while $\sigma_W$ increases from 0.3 to 2.0.
The kurtosis $\eta_W$ of $P(W)$ for multiplicative noise has a maximum around
$\tau \sim 10.0$ whereas that for additive noise keeps $\eta=3.0$
independently of $\tau$. 
We should note that $\mu_W$ and $\sigma_W$ may show oscillations
at $\tau \gtrsim 10$ if simulations are performed with finer meshes
(see Fig. 3(a) and (b) in Ref. \cite{Hasegawa11b}).
The $\tau_f$ dependence of $R$ calculated by Eqs. (\ref{eq:J1}) and(\ref{eq:J2}) is shown
in Fig. \ref{fig11}(d) where the equality: $R=\Delta F$ 
holds within conceivable numerical errors.
The JE is expected to hold in our system independently of the parameters
of $a$, $b$, $\tau$ and $\tau_f$.

Figure \ref{fig12} shows the temperature-dependent $P(W)$ 
for $k_B T=1.0$ (solid curve), $5.0$ (dashed curve) 
and $10.0$ (chain curve) 
when a ramp force of $\tau_f=10.0$ is applied to a system
with multiplicative noise of $(a, b, \tau)=(1, 0, 10.0)$.
With increasing the temperature, a width of WDF is increased and 
its departure from the Gaussian distribution is more significant: 
the kurtosis of $P(W)$ is $\eta_W=5.6$, 15.0 and 13.0 for $k_B T=1.0$,
$5.0$ and $10.0$, respectively.
The shape of non-Gaussian WDF with multiplicative noise
considerably depends on the temperature.

\section{Discussion}
The over-damped limit of the Markovian Langevin equation
is conventionally derived with setting $\ddot{x}=0$ in Eq. (\ref{eq:B2}).
This is, however, not the case where dissipation and diffusion constants
are state dependent as in our case.
Sancho, San Miguel and D\"{u}rr \cite{Sancho82} have developed 
an adiabatic elimination procedure to obtain an exact Langevin 
and FPEs in such a case.
In order to adopt their method \cite{Sancho82}, we rewrite Eq. (\ref{eq:B2}) as
\begin{eqnarray}
\ddot{x}(t) 
&=& -V'(x(t)) -\lambda(x(t)) \dot{x}(t)
+g(x(t))  \:\xi(t)+f(t),
\label{eq:D1}
\end{eqnarray}
with
\begin{eqnarray}
\lambda(x) &=& \gamma_0 \phi'(x)^2, 
\label{eq:D2}\\
g(x) &=& \sqrt{2 k_B T \gamma_0} \:\phi'(x).
\label{eq:D3}
\end{eqnarray}
By the adiabatic elimination in Eq. (\ref{eq:D1}) after Ref. \cite{Sancho82}, the FPE 
in the Stratonovich interpretation is given by 
\begin{eqnarray}
\frac{\partial P(x,t)}{\partial t} 
&=& \frac{\partial}{\partial x} \frac{1}{\lambda(x)}
\left[ V'(x)-f(t) +k_B T \frac{\partial }{\partial x} \right] P(x,t),
\label{eq:D4}
\end{eqnarray}
where we employ the relation: $g(x)^2= 2 k_B T \lambda(x)$ derived from 
Eqs. (\ref{eq:D2}) and (\ref{eq:D3}).
The corresponding Langevin equation is given by
\cite{Sancho82} 
\begin{eqnarray}
\dot{x} &=& - \frac{[V'(x)-f(t)]}{\lambda(x)}
-\frac{1}{2 \lambda(x)^2} \:g'(x) g(x) +\frac{g(x)}{\lambda(x)} \:\xi(t).
\label{eq:D5}
\end{eqnarray}
Note that the second term of Eq. (\ref{eq:D5}) does not appear when we obtain the
over-damped Langevin equation by simply setting $\ddot{x}=0$ in Eq. (\ref{eq:D1}). 
It is easy to see that the stationary distribution $P_{s}(x)$ of Eq. (\ref{eq:D4}) 
with $f(t)=0$ is given by
\begin{eqnarray}
P_{s}(x) &\propto& e^{-\beta V(x)}.
\label{eq:D6}
\end{eqnarray}

In the case of $\phi(x)=a x^2/2+ b x$ and $f(t)=0$, 
Eqs. (\ref{eq:D2}), (\ref{eq:D3}) and (\ref{eq:D5})
lead to the Langevin equation with additive and multiplicative noises given by
\begin{eqnarray}
\dot{x} &=& - \frac{V'(x)}{\gamma_0 (b+a x)^2}
-\frac{ k_B T a}{\gamma_0 (b+a x)^3} 
+\sqrt{ \frac{2 k_B T}{\gamma_0 \: (b+a x)^2} } \:\xi(t).
\label{eq:D7}
\end{eqnarray}
In the case of additive noise only ($a=0$), Eq. (\ref{eq:D7}) becomes
\begin{eqnarray}
\dot{x} &=& - \frac{V'(x)}{\gamma_0 b^2}
+\sqrt{ \frac{2 k_B T}{\gamma_0 b^2} } \:\xi(t).
\end{eqnarray}
In the opposite case of multiplicative nose only ($b=0$), we obtain
\begin{eqnarray}
\dot{x} &=& - \frac{V'(x)}{\gamma_0 a^2 x^2}
-\frac{k_B T}{\gamma_0 a^2 x^3} 
+\sqrt{ \frac{2 k_B T}{\gamma_0 a^2 x^2} } \:\xi(t).
\end{eqnarray}

Equation (\ref{eq:D7}) is quite different from a widely-adopted phenomenological
Langevin model given by \cite{Sakaguchi01,Anteneodo03,Hasegawa07}
\begin{eqnarray}
\dot{x} &=& -V'(x) + \sqrt{2 A} \:\xi(t) + \sqrt{2 M}\: x \eta(t),
\label{eq:D8}
\end{eqnarray}
where $A$ and $M$ stand for magnitudes of additive and multiplicative noises,
respectively,
and $\xi(t)$ and $\eta(t)$ express zero-mean white noise with unit variance.
The stationary PDF obtained from the PFE for Eq. (\ref{eq:D8}) 
in the Stratonovich sense is given by
\begin{eqnarray}
\ln P(x) &=& - \int \frac{V'(x)}{(A+M x^2)}\;dx 
- \left( \frac{1}{2}\right) \ln(A+Mx^2).
\label{eq:D9}
\end{eqnarray}
Equation (\ref{eq:D9}) yields Gaussian or non-Gaussian PDF for $V'(x)=x$,
depending on $A$ and $M$ \cite{Sakaguchi01,Anteneodo03,Hasegawa07}.
The Langevin model given by Eq. (\ref{eq:D8}) is one of origins leading 
to Tsallis's nonextensive statistics \cite{Tsallis}.

\section{Conclusion}
Dynamical responses and the JE of classical open systems have been studied
with the use of the generalized CL model yielding 
the non-Markovian Langevin equation in which nonlinear dissipation term
and state-dependent diffusion term satisfy the FDR [Eq. (\ref{eq:A7})]. 
Simulation results for harmonic oscillator systems are summarized as follows:

\noindent
(i) marginal stationary PDFs for $x$ and $p$ are given by $P(x)\propto e^{-\beta V(x)}$ 
and $P(p)\propto e^{-\beta p^2/2}$, respectively, independently 
of $a$, $b$ and $\tau$ (Fig. \ref{fig1}),

\noindent
(ii) $I_x(\omega_0)$ with multiplicative noise for an applied sinusoidal force 
has a peculiar $\lambda$-type peak at $\omega_0 = \omega_r$ ($> \omega_s$) 
while that with additive noise has an almost symmetric
resonance peak at $\omega_0 = \omega_s$ (Fig. \ref{fig3}), 

\noindent
(iii) enhanced fluctuations of $\rho_x(t)$ may be induced by applied 
sinusoidal force with multiplicative noise (Fig. 5), and
the $\omega_0$ dependence of its stationary $\rho_{xs}$ with multiplicative noise
has a larger $\lambda$-type peak at $\omega_0 \sim \omega_r$ for a larger $\tau$  
while $\rho_{xs}$ with additive noise is independent of
$\omega_0$ (and $\tau$) (Figs. \ref{fig6} and  \ref{fig7}), 

\noindent
(iv) dynamical responses to applied step forces considerably 
depend on $a$, $b$ and $\tau$ (Figs. \ref{fig8} and \ref{fig9}), and

\noindent
(v) the JE is valid independently of the parameters of $a$, $b$, $\tau$ and $\tau_f$
with the WDF which is Gaussian and asymmetric non-Gaussian for 
additive and multiplicative noises, respectively (Figs. \ref{fig10} and \ref{fig12}).

\noindent
The items (i)-(v) imply that a nonlinear dissipative term and a state-dependent 
diffusion term have significant effects on nonequilibrium  properties,
although they have no effects on stationary marginal PDFs.
The item (i) is in consistent with Refs. \cite{Jarzynski97,Jarzynski04},
numerically confirming the analytical result of Ref. \cite{Pollak93}.
The item (i) is, however, in contrast to the result of a widely adopted 
phenomenological Langevin model subjected to additive and multiplicative noises 
[{\it e.g.,} Eq. (\ref{eq:D8})], in which stationary PDFs depend 
on magnitudes of the two noises.
An asymmetric non-Gaussian WDF in the item (v) is similar to that in the system 
with anharmonic potential where the JE holds \cite{Mai07}. 
Theoretical results in this paper are expected to bear valuable consequences 
on the field of open systems with non-local dissipation and state-dependent diffusion.  
It would be possible to apply the present approach to classical open systems such
as free particles, anharmonic oscillators and bistable ones, 
and to extend it to quantum open systems. 


\begin{acknowledgments}
This work is partly supported by
a Grant-in-Aid for Scientific Research from 
Ministry of Education, Culture, Sports, Science and Technology of Japan.  
\end{acknowledgments}

\appendix*

\section{A. Free energy difference of the system}
\renewcommand{\theequation}{A\arabic{equation}}
\setcounter{equation}{0}
We will calculate the free energy difference of $\Delta F$, 
evaluating the system partition function $Z_S(g)$ given by Eq. (\ref{eq:J5}),
\begin{eqnarray}
Z_S(g) &=& \frac{Z(g)}{Z_B},
\label{eq:Z4}
\end{eqnarray}
where
\begin{eqnarray}
Z_B &=& {\rm Tr}\:\{e^{-\beta H_B}\}, 
\label{eq:Z5}\\
Z(g)&=& {\rm Tr}\:\{e^{-\beta[H_S(g)+H_B+H_I]} \}, 
\label{eq:Z6}
\end{eqnarray}
with
\begin{eqnarray}
H_B &=& \sum_{n=1}^N \left( \frac{p_n^2}{2 m_n}
+ \frac{m_n \omega_n^2 q_n^2}{2} \right), \\
H_S(g) &=& \frac{p^2}{2}+ \frac{\omega_s^2 x^2}{2} - x g, \\ 
&=& \frac{p^2}{2} +\frac{\omega_s^2}{2}\left(x- \frac{g}{\omega_s^2}\right)^2 
- \frac{g^2}{2 \omega_s^2}, \\
H_B + H_I &=& \sum_{n=1}^N \left[ \frac{p_n^2}{2 m_n}
+ \frac{m_n \omega_n^2 }{2}\left(q_n -\frac{c_n \phi(x)}{m_n \omega_n^2} \right)^2 \right],
\end{eqnarray}
$H_S(g)$ denoting the system Hamiltonian given by Eq. (\ref{eq:A1b}) with $f(t)=g$.
By using the Gaussian integral, we obtain the classical partition function $Z_B$
in Eq. (\ref{eq:Z5}) for $N$-body uncoupled harmonic oscillators given by
\begin{eqnarray}
Z_B &=& \frac{1}{h^N} \prod_{n=1}^{N} 
\int_{-\infty}^{\infty}  dp_n \:\int_{-\infty}^{\infty} dq_n
\: e^{-\beta (p_n^2/2m_n+ m_n \omega_n^2 q_n^2/2)}, \\
&=& \prod_{n=1}^{N} \left( \frac{2 \pi}{\beta \omega_n} \right),
\label{eq:Z2}
\end{eqnarray}
where $h$ is the Plank constant \cite{Note4}.
Similarly $Z(g)$ in Eq. (\ref{eq:Z6}) is given by
\begin{eqnarray}
Z(g) &=& \frac{e^{\beta g^2/2 \omega_s^2}}{h^{N+1}}
\int_{-\infty}^{\infty} dp \int_{-\infty}^{\infty}  dx
\: e^{-\beta [p^2/2+ (\omega_s^2/2)(x-g/\omega_s^2)^2]}  \nonumber \\
&\times& \;\prod_{n=1}^{N} \int_{-\infty}^{\infty} dp_n   
\:\int_{-\infty}^{\infty} dq_n
\: e^{-\beta \{p_n^2/2m_n+ (m_n \omega_n^2/2)[q_n-c_n \phi(x)/m_n \omega_n^2]^2\} }, \\
&=& \frac{e^{\beta g^2/2 \omega_s^2}}{h^{N+1}}
\int_{-\infty}^{\infty} dp \:\int_{-\infty}^{\infty} dy
\: e^{-\beta (p^2/2+ \omega_s^2 y^2/2)} \nonumber \\
&\times&\prod_{n=1}^{N} 
\int_{-\infty}^{\infty} dp_n  \:\int_{-\infty}^{\infty} du_n
\: e^{-\beta (p_n^2/2m_n+ m_n \omega_n^2 u_n^2/2 ) }, \\
&=& e^{\beta g^2/2 \omega_s^2} \left( \frac{2 \pi}{\beta \omega_s} \right) 
\prod_{n=1}^{N} \left( \frac{2 \pi}{\beta \omega_n} \right),
\label{eq:Z3} 
\end{eqnarray}
where we adopt changes of variables: 
$y=x-g/\omega_s^2$ and $u_n= q_n-c_n \phi(x)/m_n \omega_n^2$.
Equations (\ref{eq:Z4}), (\ref{eq:Z2}) and (\ref{eq:Z3}) lead to
\begin{eqnarray}
Z_S(g) &=& \left( \frac{2 \pi}{\beta \omega_s} \right) e^{\beta g^2/2 \omega_s^2},
\label{eq:Z1}
\end{eqnarray}
which is independent of $a$, $b$ and $\tau$.
From Eqs. (\ref{eq:J4}) and (\ref{eq:Z1}), we finally obtain
\begin{eqnarray}
\Delta F=-\frac{1}{\beta} \ln \left(\frac{Z_S(g)}{Z_S(0)} \right)
=- \frac{g^2}{2 \omega_s^2},
\label{eq:Z7}
\end{eqnarray}
which is given by Eq. (\ref{eq:J8}).
Equation (\ref{eq:Z7}) is the same as that for a linear coupling of $\phi(x)=x$
in Ref. \cite{Hasegawa11b}.


\newpage
\begin{figure}
\begin{center}
\end{center}
\caption{
(Color online) 
Marginal PDFs of (a) $P(x)$, (b) $P(p)$, (c) $P(u)$ and (d) $P(\zeta)$
for $(a,b,\tau)=(1,0,10.0)$ (solid curves), $(1,0,0.1)$ (dashed curves),
$(0,1,10.0)$ (chain curves), and $(0,1,0.1)$ (dotted curves) calculated 
by simulations with $\omega_s=1.0$, $\gamma_0=1.0$ and $k_B T=1.0$.
$P(x)$ and $P(p)$ in (a) and (b) are indistinguishable 
among the four sets of $(a, b, \tau)$.
$P(u)$ and $P(\zeta)$ for $(a,b,\tau)=(1,0,0.1)$ and (0,1,0.1) are
multiplied by a factor of five in (c) and (d) 
where PDFs are arbitrarily shifted for a clarity of figures.
}
\label{fig1}
\end{figure}

\begin{figure}
\begin{center}
\end{center}
\caption{
(Color online) 
Time courses of $\mu_x(t)$ for (a) $\tau=0.1$ and (b) $\tau=10.0$ 
with $(a, b)=(1,0)$ and $(0,1)$ when sinusoidal forces $f(t)$ with $g=1.0$ and $T_0=10.0$
shown by bottom curves are applied.
Results for $(a, b)=(0, 1)$ and $(1, 0)$ are shifted by five and ten, respectively,
for a clarity of figures.
}
\label{fig2}
\end{figure}

\begin{figure}
\begin{center}
\end{center}
\caption{
(Color online) 
The $\omega_0$ dependence of the output magnitude of $I_x(\omega_0)$ for sinusoidal forces 
with (a) $\tau=10$ and (b) $\tau=0.1$ obtained by simulations (Sim.)
for additive noise of $(a, b)=(0,1)$ (chain curves) 
and multiplicative noise of $(a, b)=(1,0)$ (solid curves),
dashed curves denoting 
theoretical (Th.) results for additive noise [Eq. (\ref{eq:C7})].
Note that ordinates of (a) and (b) are in the logarithmic and normal scales,
respectively.
}
\label{fig3}
\end{figure}

\begin{figure}
\begin{center}
\end{center}
\caption{
Lissajous plots of $\mu_x(t)$ versus $f(t)$ for (a) $T_0 = 10.0$,
(b) $T_0=5.0$, (c) $T_0=4.0$ and (d) $T_0=3.0$ with $(a, b, \tau)=(1, 0, 10.0)$.
}
\label{fig4}
\end{figure}

\begin{figure}
\begin{center}
\end{center}
\caption{
(Color online) 
Time courses of fluctuation of $\rho_x(t)$  
for sinusoidal forces with $T_0=6.0$, 10.0 and 100.0
for (a) $\tau=0.1$ and (b) $\tau=10.0$
with multiplicative noise of $(a, b)=(1, 0)$,
results being successively shifted by two for a clarity of figures.
}
\label{fig5}
\end{figure}

\begin{figure}
\begin{center}
\end{center}
\caption{
(Color online) 
The $\omega_0$ dependence of the stationary fluctuation of $\rho_{xs}$ 
for sinusoidal forces with (a) $\tau=10.0$ and (b) $\tau=0.1$
for multiplicative noise of $(a, b)=(1, 0)$ (solid curves)
and additive noise of $(a, b)=(0, 1)$ (dashed curves),
error bars expressing variations in $\rho_{xs}$.
For additive noise, $\rho_{xs}=1.0$ independently of $\omega_0$ and $\tau$.
}
\label{fig6}
\end{figure}

\begin{figure}
\begin{center}
\end{center}
\caption{
(Color online) 
The $\tau$ dependence of the stationary fluctuation of $\rho_{xs}$ 
for sinusoidal forces with $T_0=5$ (solid curve) and $T_0=10.0$ (chain curve)
for multiplicative noise of $(a, b)=(1, 0)$ and 
additive noise of $(a, b)=(0, 1)$ (dashed curve),
error bars expressing variations in $\rho_{xs}$.
}
\label{fig7}
\end{figure}

\begin{figure}
\begin{center}
\end{center}
\caption{
Time courses of $\mu_x(t)$ for (a) additive noise of 
$(a, b)=(0, 1)$ and (b) multiplicative noise of 
$(a, b)=(1, 0)$ with $\tau=0.1$, 1.0 and 10.0, 
an applied step force being plotted at bottoms.
}
\label{fig8}
\end{figure}

\begin{figure}
\begin{center}
\end{center}
\caption{
Time courses of $\mu_x(t)$ for four sets of 
$(a, b)=(0,1)$, $(0.2, 0.8)$, $(0.5, 0.5)$ and $(1,0)$ 
with $\tau=10.0$, an applied step force being plotted at the bottom.
}
\label{fig9}
\end{figure}

\begin{figure}
\begin{center}
\end{center}
\caption{
(Color online) 
WDFs of $P(W)$ for ramp forces of (a) $\tau_f=100.0$, (b) $\tau_f=10.0$ 
and (c) $\tau_f=1.0$ with $(a,b,\tau)=(1,0,10.0)$ (solid curves), 
$(1,0,0.1)$ (dashed curves), $(0,1,10.0)$ (chain curves), and $(0,1,0.1)$ (dotted curves).
Four WDFs in (c) are indistinguishable.
}
\label{fig10}
\end{figure}

\begin{figure}
\begin{center}
\end{center}
\caption{
(Color online) 
The $\tau_f$ dependence of (a) $\mu_W$, (b) $\sigma_W$, (c) $\eta_W$ and (d) $R$
for $(a,b,\tau)=(1,0,10.0)$ (solid curves), $(1,0,0.1)$ (dashed curves),
$(0,1,10.0)$ (chain curves) and $(0,1,0.1)$ (dotted curves).
Marks denote simulation results and curves are plotted only for a guide of the eye.
The arrow along the right-hand side ordinate in (d) expresses $\Delta F$ $(=- 2.0)$:
$R=\Delta F$ when the JE holds.
}
\label{fig11}
\end{figure}

\begin{figure}
\begin{center}
\end{center}
\caption{
(Color online) 
WDFs of $P(W)$ with $k_B T=1.0$ (solid curve), 5.0 (dashed curve) and
10.0 (chain curve) for a ramp force with $\tau_f=10.0$
for multiplicative noise of $(a, b, \tau)=(1, 0, 10.0)$.
}
\label{fig12}
\end{figure}


\begin{thebibliography}{99}

\bibitem{Lindner04}B. Lindner, J. Garcia-Ojalvo, A. Neiman, and L. Schimansky-Geilier,
Phys. Rep. {\bf 392}, 321 (2004).

\bibitem{Munoz04}M. A. Mu$\tilde{\rm n}$oz,
in {\it Advances in Condensed Matter and Statistical Mechanics},
ed. E. Korutcheva, R. Cuerno (Nova Publishers, New York, 2004), p. 34.

\bibitem{Sakaguchi01}H. Sakaguchi,
J. Phys. Soc. Jpn. {\bf 70}, 3247 (2001).

\bibitem{Sancho82}J. M. Sancho, M. San Miguel, and D. D\"{u}rr,
J. Stat. Phys. {\bf 28}, 291 (1982).


\bibitem{Anteneodo03}C. Anteneodo and C. Tsallis, 
J. Math. Phys. {\bf 44}, 5194 (2003).

\bibitem{Hasegawa07}H. Hasegawa,
Physica A {\bf 374}, 585 (2007).


\bibitem{Ford65}G. W. Ford, M. Kac and P. Mazur,
J. Math. Phys. {\bf 6}, 504 (1965).

\bibitem{Ullersma66}P. Ullersma,
Physica {\bf 32}, 27 (1966); {\it ibid.} {\bf 32}, 56 (1966); 
{\it ibid.} {\bf 32}, 74 (1966); {\it ibid.} {\bf 32}, 90 (1966).

\bibitem{Caldeira81}A. O. Caldeira and A. J. Leggett,
Phys. Rev. Lett. {\bf 46}, 211 (1981);
A. O. Caldeira and A. J. Leggett,
Ann. Phys. {\bf 149}, 374 (1983).


%
\bibitem{Lindenberg81}K. Lindenberg and V. Seshadri, 
Physica A {\bf 109}, 483 (1981); 
K. Lindenberg and E. Cort\'{e}s, 
ibid. {\bf 126}, 489 (1984).



\bibitem{Pollak93}E. Pollak and A. M. Berezhkovskii, 
J. Chem. Phys. {\bf 99}, 1344 (1993).





\bibitem{Barik05}D. Barik and D. S. Ray,
J. Stat. Phys. {\bf 120}, 339 (2005).

\bibitem{Chaudhuri06}J. R. Chaudhuri, D. Barik, and S. K. Banik,
Phys. Rev. E {\bf 74}, 061119 (2006).

\bibitem{Plyukhin07}A. V. Plyukhin and A. M. Froese,
Phys. Rev. E {\bf 76}, 031121 (2007).

\bibitem{Farias09}R. L. S. Farias, Rudnei O. Ramos, and L. A. da Silva,
Phys. Rev. E {\bf 80}, 031143 (2009).


\bibitem{Zaitsev09}S. Zaitsev, O. Shtempluck and E. Buks,
arXiv:0911.0833.

\bibitem{Eichler11}A. Eichler, J. Moser, J. Chaste, M. Zdrojek, I. Wilson-Rae, 
and A. Bachtold,
arXiv:1103.1788.

\bibitem{Magnasco93}M. O. Magnasco, 
Phys. Rev. Lett. {\bf 71}, 1477 (1993).

\bibitem{Julicher97}F. J\"{u}licher, A. Ajdari, and J. Prost, 
Rev. Mod. Phys. {\bf 69}, 1269 (1997).

\bibitem{Reimann02}P. Reimann, 
Phys. Reps. {361}, 57 (2002).

\bibitem{Porto00}M. Porto, M. Urbakh, and J. Klafter, 
Phys. Rev. Lett. {85}, 491 (2000); 
G. Oshanin, J. Klafter, M. Urbakh, and M. Porto, 
Europhys. Lett. {\bf 68}, 26 (2004).


\bibitem{Busta05}C. Bustamante, J. Liphardt, and F. Ritort,
Phys. Today {\bf 58}, 43 (2005).

\bibitem{Ritort07}F. Ritort,
in {\it Advances in Chemical Physics}, vol. 137, edited by S. A. Rice 
(Wiley, Hoboken, NJ, 2008) p. 31.

\bibitem{Ciliberto10}S. Ciliberto, S. Joubaud, A. Petrosyan,
J. Stat. Mech.: Theory Exp. (2010) P12003. 

\bibitem{Jarzynski97}C. Jarzynski,
Phys. Rev. Lett. {\bf 78}, 2690 (1997).

\bibitem{Evans93}D. J. Evans, E. G. D. Cohen, and G. P. Morriss,
Phys. Rev. Lett. {\bf 71}, 2401 (1993); 
D. J. Evans and D. J. Searles,
Phys. Rev. E {\bf 50}, 1645 (1994).

\bibitem{Narayan04}O. Narayan and A. Dhar,
J. Phys. A {\bf 37}, 63 (2004).

\bibitem{Crooks99}G. E. Crooks,
Phys. Rev. E {\bf 60}, 2721 (1999). 

\bibitem{Jarzynski97b}C. Jarzynski,
Phys. Rev. E {\bf 56}, 5018 (1997).

\bibitem{Jarzynski04}C. Jarzynski,
J. Stat. Mech.: Theory Exp. (2004)  P09005.

\bibitem{Liphardt02}J. Liphardt, S. Dumont, S. Smith, I. Tinoco, C. Bustamante, 
Science {\bf 296}, 1833 (2002).

\bibitem{Wang05}G. M. Wang, J. C. Reid, D. M. Carberry, D. R. M. Williams, 
E. M. Sevick, and Denis J. Evans,
Phys. Rev. E {\bf 71}, 046142 (2005).

\bibitem{Douarche05}F. Douarche, S. Ciliberto, A. Petrosyan,
and I. Rabbiossi,
Europhys. Lett. {\bf 70}, 593 (2005).

\bibitem{Douarche06}F. Douarche, S. Joubaud, N. B. Garnier, 
A. Petrosyan, and S. Ciliberto, 
Phys. Rev. Lett. {\bf 97}, 140603 (2006). 

\bibitem{Joubaud07}S. Joubaud, N. B. Garnier, F. Douarche, A. Petrosyan, and
S. Ciliberto,
C. R. Physique {\bf 8}, 518 (2007).

\bibitem{Joubaud07b}
S. Joubaud, N. B. Garnier, S. Ciliberto,
J. Stat. Mech.: Theory Exp. (2007) P09018.

\bibitem{Zamponi05}F. Zamponi, F. Bonetto, L. F. Cugliandolo and J. Kurchan,  
J. Stat. Mech.: Theory Exp. (2005) P09013.

\bibitem{Mai07}T. Mai and A. Dhar,
Phys. Rev. E {\bf 75}, 061101 (2007).

\bibitem{Speck07}T. Speck and U. Seifert,
J. Stat. Mech.: Theory Exp. (2007) L09002.

\bibitem{Ohkuma07}T. Ohkuma and T. Ohta,
J. Stat. Mech.: Theory Exp. (2007) P10010.


\bibitem{Chaudhury08}S. Chaudhury, D. Chatterjee and B. J Cherayil,
J. Stat. Mech.: Theory Exp. (2008) P10006.


\bibitem{Dhar05}A. Dhar,
Phys. Rev. E {\bf 71}, 036126 (2005).

\bibitem{Jarzynski06}C. Jarzynski,
Comptes Rendus Physique {\bf 8}, 495 (2007).

\bibitem{Jarzynski08}C. Jarzynski,
Eur. Phys. J. B. {64}, 331 (2008).

\bibitem{Chakrabarti08}R. Chakrabarti,
arXiv:0802.0268.


\bibitem{Hijar10}H. Hijar and J. M. O. de Z\'{a}rate,
Eur. J. Phys. {\bf 31}, 1097 (2010).

\bibitem{Hasegawa11b}H. Hasegawa,
Phys. Rev. E {\bf 84}, 011145 (2011).

\bibitem{Saha06}A. Saha and J. K. Bhattacharjee,
J. Phys. A {\bf 40}, 13269 (2007).

\bibitem{Hasegawa11c}H. Hasegawa,
arXiv:1104.4756.


\bibitem{Lev10}B. I. Lev and A. D. Kiselev,
Phys. Rev. E {\bf 82}, 031101 (2010).

\bibitem{Aron10}C. Aron, G. Biroli, and L. F. Cugliandolo,
J. Stat. Mech.: Theory Exp. (2010) P11018. 

%
\bibitem{Bao05}J-D. Bao, Y-L. Song, Q. Ji, and Y-Z. Zhuo,
Phys. Rev. E {72}, 011113 (2005).


\bibitem{Note1}In the Heun method for the ordinary differential equation of 
$dx/dt=f(x)$, a value of $x$ at $t+h$ is evaluated by 
$x(t+h)=x(t)+(h/2)[f(x_0)+f(x_1)]$ with $x_0=x(t)$ and $x_1=x(t)+h f(x(t))$, 
while it is given by $x(t+h)=x(t)+h f(x(t))$ in the Euler method, $h$ being the time step. 
The Heum method for the stochastic ordinary differential equation meets 
the Stratonovich calculus employed in the FPE \cite{Note2}.

\bibitem{Note2}W. R\"{u}melin, SIAM J. Numer. Anal. {\bf 19}, 604 (1982);
A. Greiner, W. Strittmatter, and J. Honerkamp, J. Stat. Phys. {\bf 51} 95 (1988);
R.F. Fox, I.R. Gatland, R. Roy, and G. Vemuri, Phys. Rev. A {\bf 38} 5938 (1988).

\bibitem{Tsallis}C. Tsallis, J. Stat. Phys. {\bf 52} (1988) 479; 
C. Tsallis, R. S. Mendes and A. R. Plastino,
Physica A {\bf 261}, 534 (1998);
C. Tsallis, Physica D {\bf 193}, 3 (2004). 

\bibitem{Note4}
We may set $h=1.0$ becasue its precise value does not matter in the classical limit.


\end{thebibliography}
\end{document}